\documentclass[10pt,
							 twocolumn,
							 superscriptaddress,
							 english,
							 prl,
							 showpacs,
							 floatfix,
							 aps
							]{revtex4-1}
\usepackage[utf8]{inputenc}
\usepackage{amsmath}
\usepackage{amssymb}
\usepackage{graphicx}
\usepackage{xspace}
\usepackage{ulem}

\makeatletter
\@ifundefined{textcolor}{}
{%
 \definecolor{BLACK}{gray}{0}
 \definecolor{WHITE}{gray}{1}
 \definecolor{RED}{rgb}{1,0,0}
 \definecolor{GREEN}{rgb}{0,1,0}
 \definecolor{BLUE}{rgb}{0,0,1}
 \definecolor{CYAN}{cmyk}{1,0,0,0}
 \definecolor{MAGENTA}{cmyk}{0,1,0,0}
 \definecolor{YELLOW}{cmyk}{0,0,1,0}
}

\usepackage{soul}
\usepackage{braket}
\usepackage[backref=none,
bookmarksnumbered=true,
bookmarks=true,
bookmarksopen=true,
colorlinks=true,
citecolor=blue,
linkcolor=blue,
anchorcolor=green,
urlcolor=blue,unicode=false]{hyperref}

\renewcommand{\v}[1]{\ensuremath{\mathbf{#1}}} 
 
\let\baraccent=\= 
\renewcommand{\=}[1]{\stackrel{#1}{=}} 

\newcommand{\unitspace}{~}

\newcommand{\didv}{\ensuremath{\mathrm{d}I/\mathrm{d}V}\xspace}
\newcommand{\Figref}[1]{Figure~\ref{#1}}
\newcommand{\mos}{MoS$_2$}
\newcommand{\HPc}{H$_2$Pc}

\DeclareMathOperator{\uV}{\unitspace\mathrm{V}}
\DeclareMathOperator{\umV}{\unitspace\mathrm{mV}}

\DeclareMathOperator{\upA}{\unitspace\mathrm{pA}}

\makeatother

\usepackage{babel}

\begin{document}

\title{Vibrational excitation mechanism in tunneling spectroscopy beyond the Franck-Condon model}

\author{Ga\"el Reecht}
\affiliation{\mbox{Fachbereich Physik, Freie Universit\"at Berlin, Arnimallee 14, 14195 Berlin, Germany}}

\author{Nils Krane}
\affiliation{\mbox{Fachbereich Physik, Freie Universit\"at Berlin, Arnimallee 14, 14195 Berlin, Germany}}

\author{Christian Lotze}
\affiliation{\mbox{Fachbereich Physik, Freie Universit\"at Berlin, Arnimallee 14, 14195 Berlin, Germany}}

\author{Lei Zhang}
\author{Alejandro L. Briseno}
\affiliation{\mbox{University of Massachusetts, Department of Polymer Science and Engineering, Amherst, USA}}

\author{Katharina J. Franke}
\affiliation{\mbox{Fachbereich Physik, Freie Universit\"at Berlin, Arnimallee 14, 14195 Berlin, Germany}}

\date{\today}

\begin{abstract}
Vibronic spectra of molecules are typically described within the Franck-Condon model. Here, we show that highly resolved vibronic spectra of large organic molecules on a single layer of \mos\ on Au(111) show spatial variations in their intensities, which cannot be captured within this picture. We explain that vibrationally mediated perturbations of the molecular wave functions need to be included into the Franck-Condon model. Our simple model calculations reproduce the experimental spectra at arbitrary position of the STM tip over the molecule in great detail.

\end{abstract}

\maketitle 


Vibronic excitations are resonant transitions from a molecular ground state to an electronically and vibrationally excited state. These excitations are typically described by the Franck-Condon model. The essence of it are fast electronic transitions treated in Born-Oppenheimer approximation, such that the excitations occur without changes in the nuclear coordinates or momentum. Vibronic excitations in single molecules on surfaces can be detected as resonant sidebands of positive or negative ion resonances in tunneling spectroscopy \cite{Qiu2004, Pradhan2005, Nazin2005, Frederiksen2008,Matino2011, Schulz2013,Wickenburg2016} with apparent submolecular variations due to distinct close-lying orbitals \cite{Ogawa2007, Huan2011, Mehler2018}. In contrast to these resonant excitations, inelastic vibrational excitations far below resonance \cite{Stipe1998, Heinrich2002, Ho2002} are described by a change in the nuclear coordinates, which leads to a modified tunneling matrix element and to the opening of a new tunneling path \cite{Lorente2000, Lorente2004, Burema2012}. Hence, off-resonant inelastic tunneling and resonant vibronic transitions are treated in distinct and complementary models \cite{Franke2012}. The combination of both models would be akin to phonon-mediated electronic transitions in crystal structures \cite{Chynoweth1962,Eaves1985}, where the activation of a phonon mode enables otherwise forbidden electronic transitions, as the initial and final state have different parallel momentum in the electronic band structure \cite{Zhang2008, Wehling2008, Natterer2015, Vdovin2016}. Signatures of such combined excitations in single molecules have not been reported to date. 

Recent tunneling experiments have revealed some limitations of the Franck-Condon model. In cases, where the electronic energy level spacing was similar to vibrational energies, it was found that avoided level crossings determine the resonant sidebands \cite{Repp2010, Schwarz2015}. In other cases, intensity variations of the resonant sidebands along an organic molecule were interpreted in terms of coherent vibrational modes with different symmetries \cite{Ogawa2007} or with vibration-assisted coupling of wave functions of different symmetry in molecule and tip \cite{Pavlicek2013}. Selection rules could not be derived, because the vibrational modes and associated nuclear displacements of the molecule could not be identified owing to an insufficient experimental energy resolution probably limited by non-adiabatic relaxation effects.

Here, we show that vibration-assisted tunneling and Franck-Condon excitations are crucial for a complete vibronic model. 
To benchmark our model we use vibronic spectra of large organic molecules on a single layer of \mos\ on Au(111). The van-der-Waals layer acts as an efficient decoupling layer from the metal substrate and provides exceptional energy resolution of a few meV \cite{Krane2018, Krane2019, Reecht2019}. This allows us to probe vibronic states and their modulation of intensities with intramolecular resolution. We show that the spatial intensity variations can be simulated by including vibration-assisted tunneling in addition to the Franck-Condon picture.

Scanning tunneling microscopy (STM) experiments were performed at a temperature of 4.6\,K in ultra-high vacuum. 
Monolayer-islands of \mos\, were grown on a clean Au(111) surface by depositing Mo in an H$_2$S atmosphere ($5\cdot 10^{-5}$ mbar) and annealing to 800\,K \cite{Gronborg2015,Krane2016}. 2,5-Bis\-(3-dodecyl\-thiophen-2-yl)thieno[3,2-b]thiophene (BTTT) [phthalocyanine (\HPc)] molecules were evaporated at 365\,K [680\,K] onto the surface held at 200\,K [120\,K]. Differential-conductance (\didv) spectra and maps were recorded using lock-in detection with 921\,Hz modulation frequency.

\begin{figure}[tb]
	\includegraphics[width=\columnwidth]{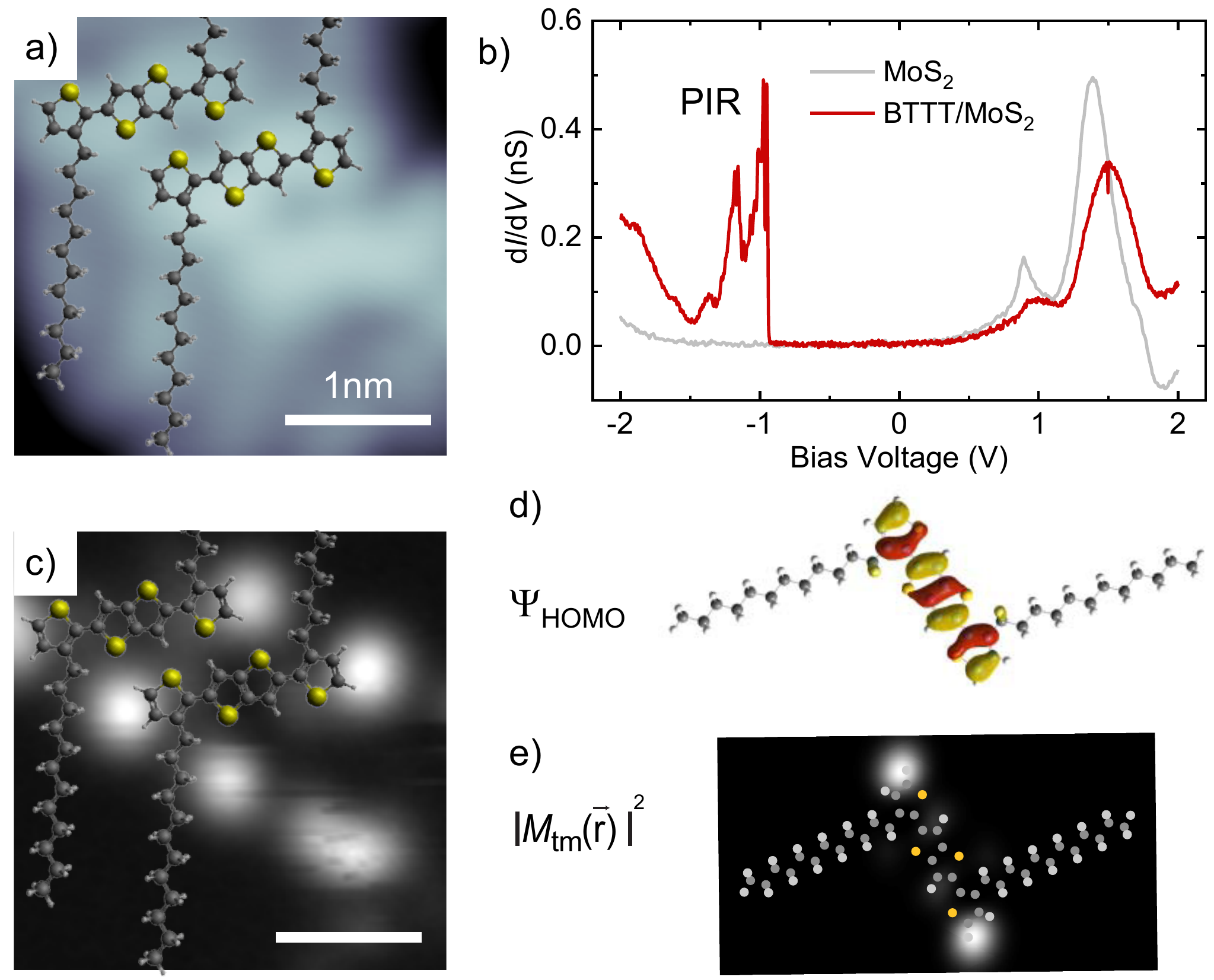}
	\caption{a) STM topography of BTTT on \mos/Au(111) with overlaid molecular model (I=20$\upA$,V=$1\uV$). b) \didv spectra recorded over bare \mos\,(gray) and BTTT (red) (feedback opened at: 150$\upA$, $2.2\uV$; V$_\mathrm{mod}$=10$\umV$). c) Constant-height \didv map of the same area as a) at V = -0.975\,$\uV$ (V$_\mathrm{mod}$=10$\umV$). d) DFT-calculated HOMO iso-density ($\Psi_\mathrm{HOMO}$) of BTTT and e) position-dependent calculation of the tunneling matrix element $\left|M_{tm}(\vec{r}) \right|^2$ between $\Psi_\mathrm{HOMO}$ and an s-type wave function $\Psi_{t}$ of the STM tip (tip-molecule distance 8.5\,\AA \cite{note}). }
	\label{fig1}
\end{figure} 

Deposition of BTTT molecules on \mos\, leads to partially ordered structures with the molecules lying parallel to each other (\Figref{fig1}a) \cite{note2}. \didv spectra recorded at the extremity of a BTTT (red curve) and on the bare \mos\, (gray curve) are presented in \Figref{fig1}b. The spectrum on \mos\,  shows the well characterized semiconducting band gap ($\sim$\,-1.4\,V to +0.5\,V)\cite{Bruix2016}, which is essential for decoupling the molecules from the substrate \cite{Krane2018, Reecht2019}. The spectrum on the BTTT molecules shows a positive ion resonance inside this gap at $\sim$\,-1\,V. This resonance exhibits a rich satellite structure revealing the vibronic properties of the molecule. The intensity of this resonance is largest at the ends of the thiophene backbone (\didv map in \Figref{fig1}c). At first sight this shape does not agree with the delocalized nature of the highest occupied molecular orbital (HOMO) (\Figref{fig1}d). 
However, we note that the conductance signal in STM is proportional to the square of the tunneling matrix element $M_{tm}$, which depends on the overlap of the wave functions $\Psi_t$ of the tip and $\Psi_m$ of the molecule \cite{Bardeen1961}.

We simulate the position-dependent tunneling matrix element along the BTTT molecule by assuming an s-type wave function for the STM tip and the DFT-derived molecular wave function (details in supplemental material (SM)). The simulation for the HOMO (\Figref{fig1}e) is in good agreement with the constant-height \didv map of the positive ion resonance inside the gap (\Figref{fig1}c). 

\begin{figure}[tb]
	\includegraphics[width=\columnwidth]{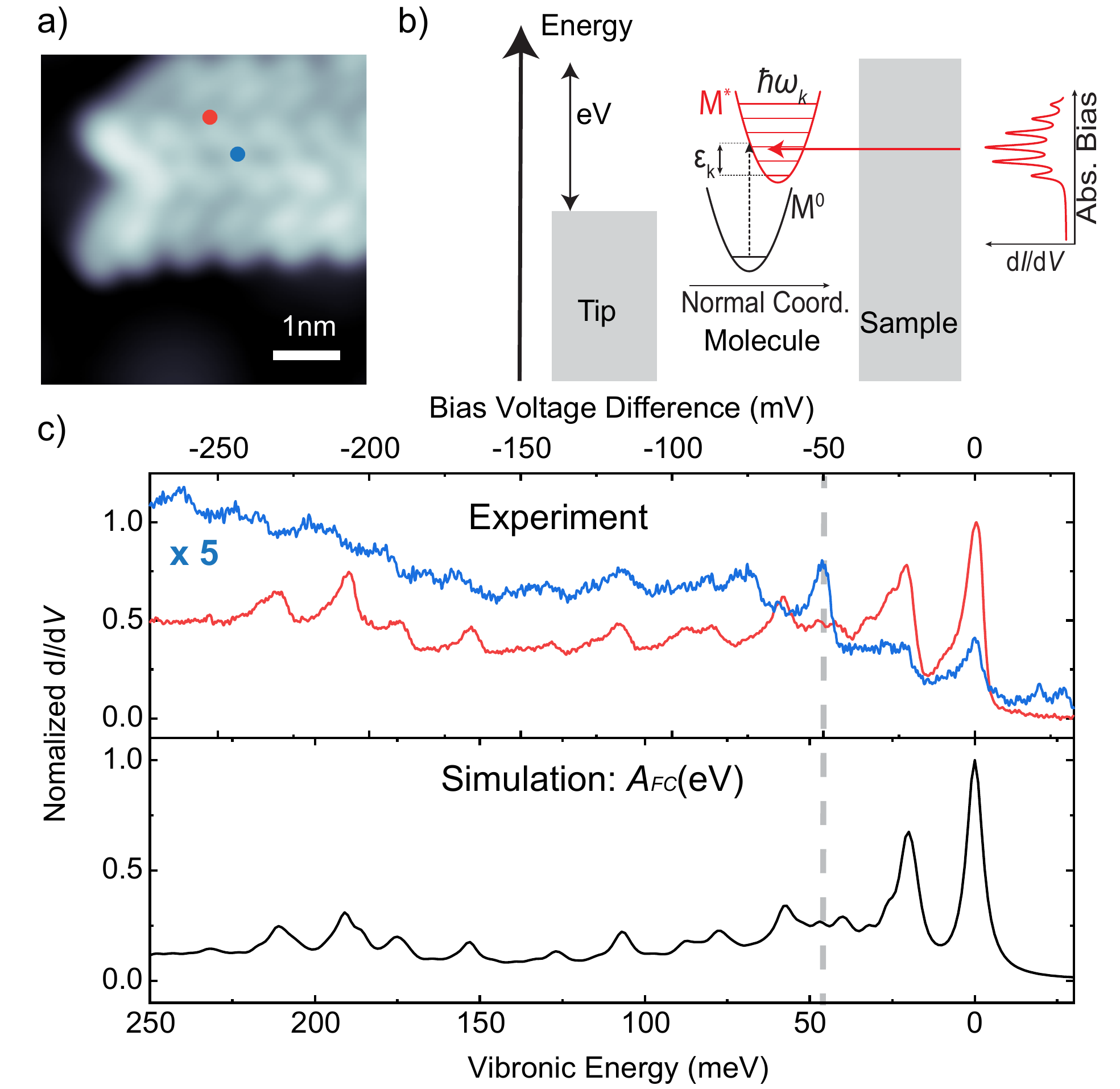}
	\caption{a) STM topography of BTTT on \mos/Au(111) (I=10$\upA$, V=0.9$\uV$). Sketch of  vibrational excitation mechanism of a molecule in an STM junction. c) Top: \didv spectra at the PIR of BTTT recorded at the same tip height on two different positions of the same molecule (see corresponding dots in a)). Spectra are normalized to the largest intensity of the red spectrum. The blue spectrum is then enlarged by a factor of 5 and shifted in energy to the same elastic peak (feedback opened at: 250$\upA$, -1.2$\uV$ at extremity, V$_\mathrm{mod}$=0.5$\umV$). Bottom: simulated \didv vibronic spectrum considering the Franck-Condon mechanism described in b), energy broadening set to FWHM=6\,mV. The upper and lower axis differ by $10\%$ to account for the voltage drop in \mos.}
	\label{fig2}
\end{figure} 	

We now focus on the study of the vibrational properties of BTTT. Highly-resolved spectra of the satellite structure of the PIR recorded at the extremity (red) and in the center (blue) of the thiophene backbone are shown in \Figref{fig2}c. Both spectra are normalized to the highest-intensity peak of the red spectrum, and a factor 5 is additionally applied to the blue one for clarity. Furthermore, they are shifted to the same onset energy, \textit{i.e.}, to the elastic peak, to eliminate the effect of the inhomogeneous tip potential over the molecule \cite{Krane2019} (see raw data in SM). We first discuss the spectrum recorded at the extremity of the molecule, corresponding to the position of highest tunneling probability into the HOMO-derived resonance (see \Figref{fig1}c). As described in a previous work on BTTT, the vibronic fingerprint at this site can be explained within the Franck-Condon picture \cite{Krane2018}. The excitation probability of a vibrational mode $k$ and its harmonics is given by the overlap integral of the initial (ground state) and final state. It depends on the relaxation energy $\epsilon_k$ when charging the molecule,  as sketched in \Figref{fig2}b. The relaxation energy of each vibrational mode can be derived from gas-phase DFT calculations of the neutral and the positively charged molecular state (details in SM). A simulated Franck-Condon spectrum $A_\mathrm{FC} (eV)$ for the BTTT molecule is shown in the bottom panel of \Figref{fig2}c. It is in remarkable agreement with the experimental spectrum recorded at the extremity of the molecule (red). Note that a scaling factor is required between the calculated vibronic energy (bottom axis) and the bias voltage of experimental \didv spectra (top axis), accounting for the voltage drop in the \mos\, layer ($\sim 10\%$). 

Considering only the Franck-Condon principle, the relative intensities of the vibronic resonances should be constant along the molecule. This is in contrast to the experimentally observed spatial variations of excitation efficiencies along the molecule in \Figref{fig2}c. The spectrum at the center of the BTTT molecule exhibits different intensity ratios of the vibronic peaks from the spectrum at the extremities. Most striking is a strongly enhanced peak at $\sim 50$\,meV above the elastic peak (dashed line)  \cite{note3}. 
We will show that this additional peak can be explained by an additional excitation mechanism.

\begin{figure}[tb]
	\includegraphics[width=\columnwidth]{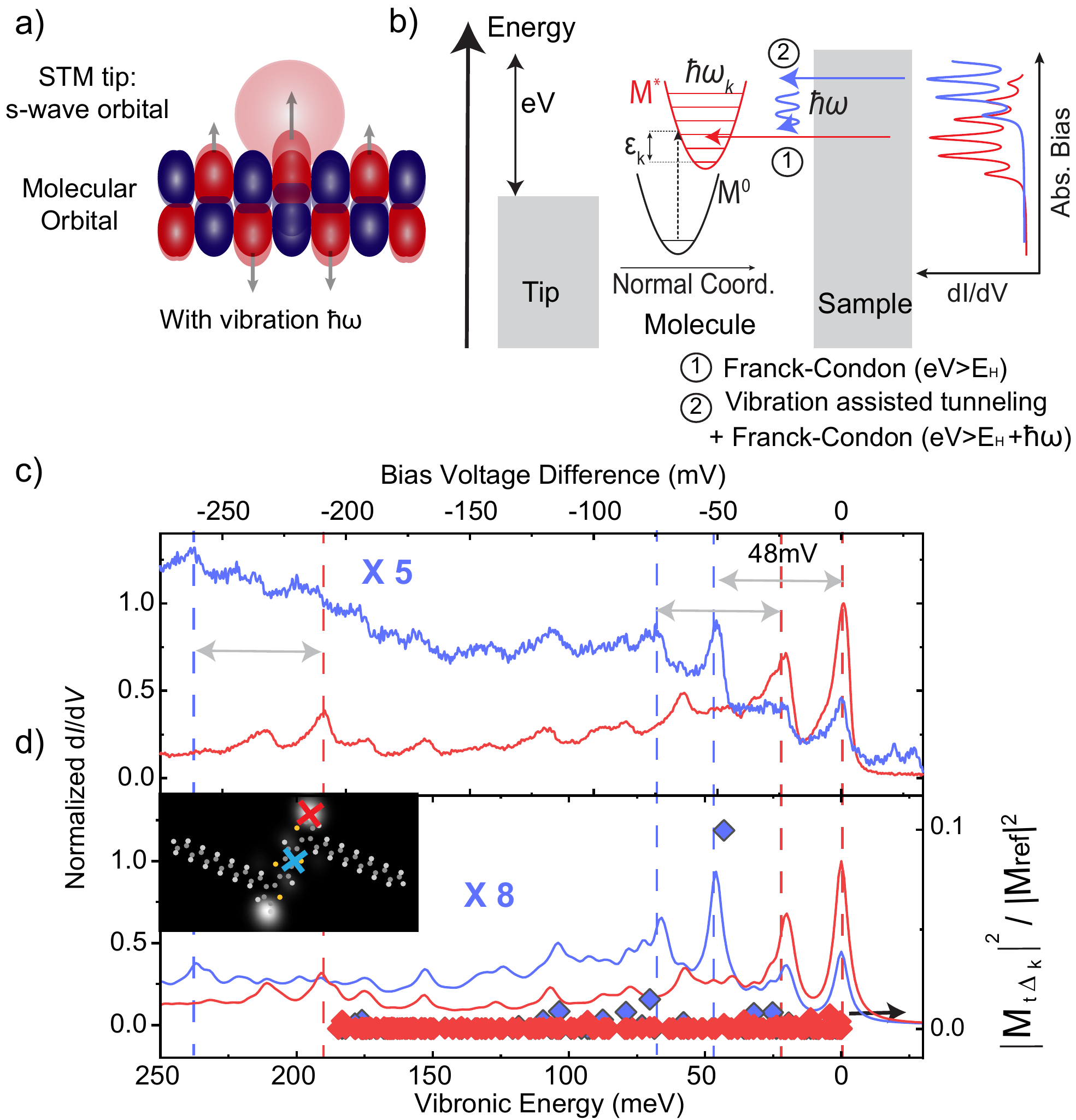}
	\caption{a) Schematic representation of vibration-assisted tunneling: when the tip is placed over the center of the molecule, the overlap between the tip and molecule wave function is small due to the nodal planes. Considering a vibrational distortion of the molecule the overlap can be modified. If the overlap is increased, the tunneling matrix element is increased and a new inelastic channel is opened, as sketched in b). c) Experimental spectra of Fig. \ref{fig2}c of the PIR at two different tip positions (center: blue; extremity: red). d) Diamonds (right axis): calculated change of the tunneling matrix element $\left|M_{t\Delta_k}(\vec{r})\right|^2$ for all vibrational modes $k$ of BTTT, for tip position similar to the experimental spectra in b) (See map and model with crosses in inset, tip height 8.5\,\AA). $\left|M_{t\Delta_k}(\vec{r})\right|^2$ are given in units of $M_\mathrm{ref}$ being the maximum value of $M_{tm}$, $i.e.$, at the extremity of the molecule. Line (left axis): simulated spectra for the same tip positions including vibration-assisted tunneling using equation \ref{Sp_tot}.}
	\label{fig3}
\end{figure} 

Pavlicek \textit{et al.} \cite{Pavlicek2013} have recently explained spatial variations of vibronic excitations by the coupling of wave functions of different symmetry in the tip and molecule. However, the vibronic peaks were of several tens of meV width probably due to non-adiabatic coupling to substrate phonons in the ionic NaCl substrate \cite{Repp2005}. The lack of energy resolution thus prevented the identification of the involved vibrational modes. Consequently, the determination of selection rules of this excitation was not possible.           
To explain the spatial variation of the vibronic signature across the BTTT molecule, we propose the following mechanism.
In addition to the Franck-Condon picture (process 1 in \Figref{fig3}b) we also consider vibration-assisted tunneling (VAT; process 2 in \Figref{fig3}b).
The latter is in analogy to the off-resonant inelastic excitation, also referred to as inelastic electron tunneling spectroscopy (IETS).
As discussed before, the tunneling probability into a molecular orbital (MO)  is determined by the overlap of the tip and MO electronic wave function.
This overlap is strongly reduced for an s-wave tip at the center of the BTTT, due to the symmetry of the HOMO's nodal planes (see Figure~1c,e and scheme in  Figure~3a).
In VAT, this symmetry can be broken by a vibrational mode of the molecule as shown in \Figref{fig3}a.
Here, an out-of-plane mode with an anti-symmetric character relative to the HOMO's nodal planes leads to an increasing (decreasing) overlap between tip and the parts of the molecular wave function with positive (negative) sign, or vice versa. Such a change of the wave function allows for a large (vibration-assisted) tunneling matrix element  \cite{Lorente2000, Schwarz2015} and effectively opens a new tunneling channel at the threshold of the vibrational excitation, \textit{i.e.} at $E_V=E_H + \hbar\omega_k$ with $E_H$ being the elastic excitation of the HOMO (see scheme in \Figref{fig3}a-b). In contrast to off-resonant inelastic tunneling, resonant VAT includes an electronic excitation according to the Franck-Condon principle and an additional inelastic transition within the excited state
This combination leads to a replicas of the Franck-Condon spectrum at higher energies (\Figref{fig3}b).

To analyze which vibrations give rise to a large VAT matrix element, we assume -- in analogy to off-resonant inelastic excitation --  the tunneling probability of the newly opened channel to depend on the overlap between s-wave tip and perturbed molecular wave function $\delta Q_k \frac{\partial \Psi_m}{\partial Q_k}$ \cite{Lorente2000,Burema2012}. In our calculation we approximate it by:
\begin{equation}
 \delta Q_k \frac{\partial \Psi_m}{\partial Q_k} \approx \delta Q_k \frac{\Delta \Psi_m}{\Delta Q_k} = \frac{\Psi_{m,k+} - \Psi_{m,k-}}{2}=\frac{\Delta\Psi_{m,k}}{2},
	\label{eq:Approx}
\end{equation}
with $\Psi_{m,k\pm}$ being the wave function for the vibrationally excited molecule, displaced by $\pm \delta Q_k$ as determined from DFT calculations of the free molecule (details in SM). Thus, we calculate the VAT matrix element $\left|M_{t\Delta_k}(\vec{r})\right|^2$ (see equation S6 in SM) for the perturbed molecular wave function at the center and extremity of the BTTT molecule  for each mode [\Figref{fig3}d (diamonds)]. At the center of the molecule, one vibrational mode at $43.1$\,meV  yields a large change of the tunneling matrix element. Considering the voltage drop in the \mos\, layer ($\sim 10\%$), this is in good agreement with the additional peak at $\sim 50$\,meV in the experimental \didv spectrum (blue spectrum). This mode corresponds to an out-of-plane mode, mainly localized on the thienothiophene unit (see SM). In contrast, at the molecular extremity the change in tunneling matrix element is negligible for all modes. 
Thus, spectra taken at this position can be explained by considering only the Franck-Condon principle.

In order to merge the two excitation mechanisms, we consider that the VAT adds another tunneling channel to the excited molecule (in its PIR) of intensity $\left|M_{t\Delta_k}(\vec{r})\right|^2$, which then has to be convolved with the Franck-Condon spectrum $A_\mathrm{FC}(eV)$. As a result, the Franck-Condon peaks are replicated at energies shifted by $\hbar\omega_k$. A fingerprint of this can be seen by the repetition of the pattern of the most intense peaks indicated by red and blue dashed lines in \Figref{fig3}c. Thus, the position-dependent vibronic intensities can be calculated as (details in SM): 

\begin{equation}
\begin{aligned}
A_\mathrm{total}(\vec{r},eV) = \left|M_{tm}(\vec{r})\right|^2 \cdot A_\mathrm{FC}(eV) \\
+  \sum_{k} \left|M_{t\Delta_k}(\vec{r})\right|^2 \cdot A_\mathrm{FC}(eV+\hbar \omega_k),
\end{aligned}
	\label{Sp_tot}
\end{equation}
      
Note that the only free parameter for the simulation is the tip-molecule distance (discussion in SM). \Figref{fig3}d shows the simulated spectra for the tip over the center (blue) and extremity (red) of the molecule for a tip--molecule distance of 8.5\,\AA\ \cite{note}. Both are now in good agreement with the experimental spectra. Equivalent simulations can be carried out at arbitrary position over the molecule (see SM).

\begin{figure}[tb]
	\includegraphics[width=\columnwidth]{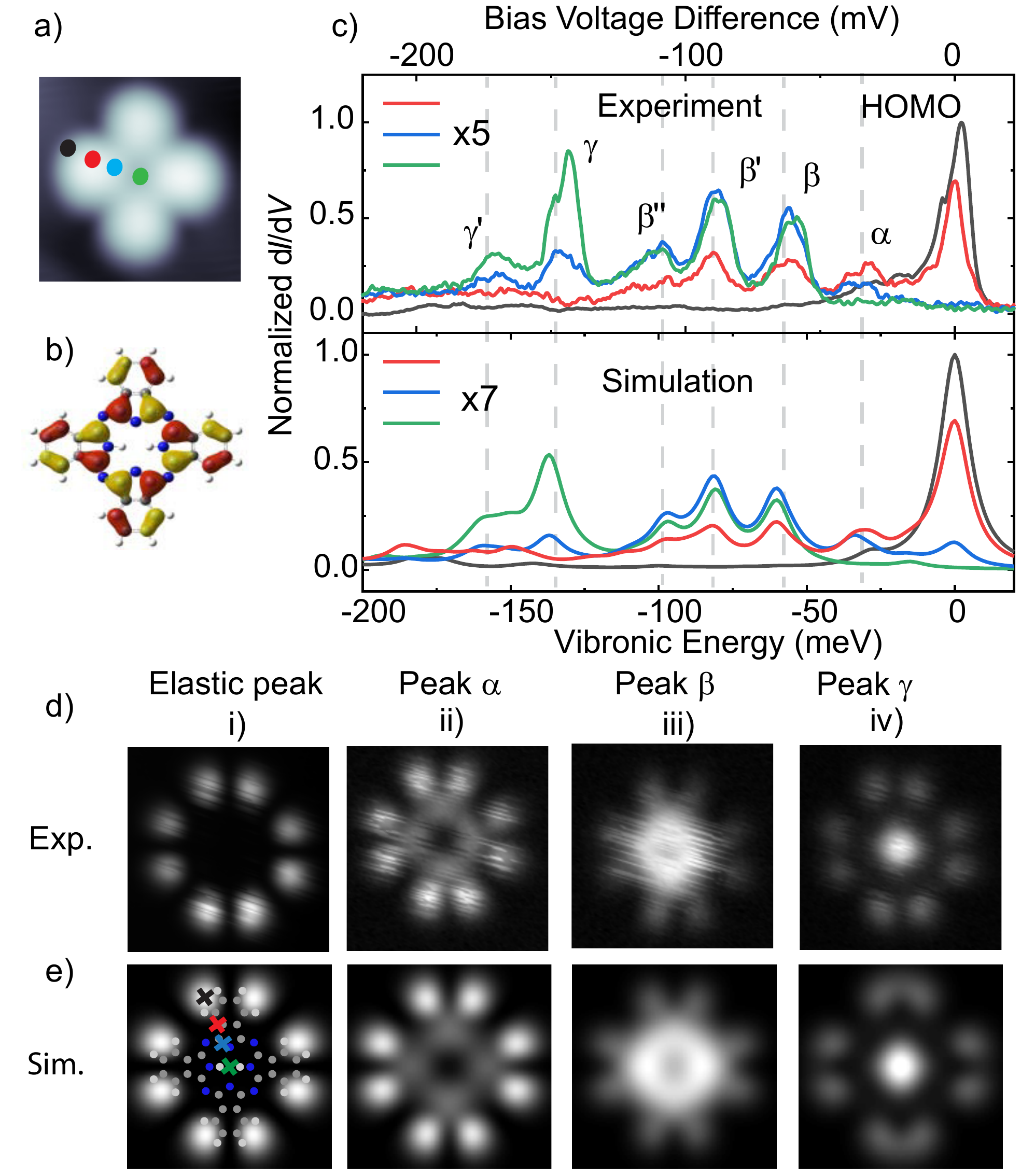}
	\caption{a) 2.4 x 2.4\,nm$^2$ STM topography of \HPc\, on \mos/Au(111) (I=20$\upA$,V=0.5$\uV$). b) DFT-calculated HOMO iso-density of \HPc. c) Top: \didv spectra recorded at the same tip height at positions shown by dots in a) (feedback opened at: 40$\upA$, -1.2$\uV$ on lobe of \HPc,  V$_\mathrm{mod}$=2$\umV$). Spectra are normalized to the elastic peak of the black spectrum. Additionally, the red, blue and green spectra are scaled by a factor 5 for clarity. All spectra are shifted to the same energy of the elastic peak. Bottom: simulated spectra for tip positions indicated as crosses in e), tip height 7.5\,\AA).  d)  2.4 x 2.4\,nm$^2$ constant-height \didv map at energies corresponding to (i) the elastic peak (HOMO) and (ii-iv) $\alpha$, $\beta$ and $\gamma$ inelastic peaks in b) (V$_\mathrm{mod}$=5$\umV$). e) Corresponding simulated conductance maps. }
	\label{fig4}
\end{figure}

To validate our full vibronic model, we now turn to a second molecular system: phthalocyanine (\HPc). This molecule appears with a clover shape when scanned at low bias (\Figref{fig4}a) and exhibits -- as BTTT -- a sharp PIR in the semiconducting \mos\, gap \cite{Reecht2019} (see large-scale spectra in SM), associated to the removal of one electron from its HOMO. The \didv map recorded at the energy of the PIR shows eight lobes, two lobes next to each isoindole moiety [\Figref{fig4}d(i)]. As for BTTT, this map can be simulated [\Figref{fig4}e(i)] by considering the tunneling matrix element between an s-wave tip and the DFT-calculated HOMO wave function of  \HPc\ (\Figref{fig4}b). Highly resolved spectra of the PIR recorded over one of this eight lobes show mainly one sharp elastic peak  (black spectrum in \Figref{fig4}c). At this position, there is hardly any intensity in the vibronic sidebands. This can be understood by a small relaxation energy $\epsilon_k=21.7$\,meV of \HPc\, upon positive charging (simulated Franck-Condon spectrum in SM).

In contrast, spectra over the isoindoline moieties show a pronounced set of peaks above the elastic peak (\Figref{fig4}b). The peaks can be classified into three groups ($\alpha$,$\beta$ and $\gamma$) according to their spatial distribution in the respective \didv maps [see experimental maps in \Figref{fig4}d(ii-iv)]. While all maps show the eight-lobe structure (with varying intensity), the $\alpha$ resonances show additional intensity on the macrocycle, the $\gamma$s have very strong weight on the center and $\beta$s show some conductance on the macrocycle and the center (all maps in SM). 

To show that these position-dependent resonances result from the above derived picture of a combination of VAT and Franck-Condon excitations, we calculate the VAT matrix element of  the vibrationally perturbed molecular wave functions with an s-wave tip. We find a large numbers of  vibrational modes, which cause an increase of the VAT matrix element (see SM). The corresponding calculated \didv spectra  (according to equation \ref{Sp_tot}, bottom panel of \Figref{fig4}b) are in excellent agreement with the experimental spectra. We note that the elastic peak can be of much lower intensity than the inelastic satellites. To prove that our model reproduces the spatial distribution of the VAT, we also simulate the conductance maps at the energies of the different vibrational modes involved in the transport, which are in remarkably good agreement with the experimental maps (\Figref{fig4}e).

In conclusion, our highly-resolved tunneling spectra of molecules on a monolayer of \mos\ revealed evidence of the interplay of two distinct excitation mechanisms of molecular vibrations. On the one hand, tunneling electrons resonantly excite vibrational modes which can be described within the Franck-Condon model. On the other hand, inelastic tunneling electrons excite molecular vibrations, which facilitate tunneling into the molecule at positions, where it would be suppressed in the static case for symmetry reasons. We could describe all spectral details of vibrational intensities by combining vibration-assisted tunneling (VAT) and the Franck-Condon principle. We note that VAT may enhance higher-energy resonances over the elastic onset. In the case of less resolved spectra, this may cause an apparent shift of peaks. Furthermore, our description of VAT is not restricted to STM experiments, it may also affect charge transport through organic materials and heteromolecular interfaces.

\acknowledgements
This work was supported by the Deutsche Forschungsgemeinschaft (DFG) - Projektnummer 182087777 - SFB 951, and by the European Research Council (ERC) through the Consolidator Grant "NanoSpin". 
We gratefully acknowledge discussions with M.-L. Bocquet and A. Donarini.

\bibliographystyle{apsrev4-1}

%

\clearpage

\newcommand{\beginsupplement}{%
	\setcounter{table}{0}
	\renewcommand{\thetable}{S\arabic{table}}%
	\setcounter{figure}{0}
	\renewcommand{\thefigure}{S\arabic{figure}}%
	\setcounter{equation}{0}
	\renewcommand{\theequation}{S\arabic{equation}}%
	\setcounter{section}{1}
	\renewcommand{\thesection}{S\arabic{section}}%
	}
\onecolumngrid

\section*{\Large{Supplemental Material}}

\section{Simulation details}
\subsection{Spatially resolved tunneling matrix element}

In the main text, we compared spatially resolved maps of the square of the tunneling matrix element along the flat-lying molecule with constant-height \didv maps. Here, we present some more details of the simulations. The tunneling matrix element is given by \cite{Bardeen1961}:

\begin{equation}
	\left|M_{tm}(\vec{r}) \right|^2 = \left|\frac{\hbar^2}{2m} \int \Psi_{t} \vec{\nabla} \Psi_{m} - \Psi_{m} \vec{\nabla} \Psi_{t} d\vec{S}\right|^2,
	\label{eq:Bardeen}
\end{equation} 

with $\Psi_{m}$ being the wave function of the molecule and $\Psi_{t}$ a spherical wave function of an s-wave tip. The molecular wave functions are obtained by DFT calculations of the isolated molecules, using GAUSSIAN09 package with the B3PW91 functional and the 6-31g(d,p) basis set \cite{Gaussian}. As we observe a positive ion resonances of BTTT and \HPc\ inside the semiconducting band gap of \mos, we calculate the spatial distributions of $\left|M_{tm}(\vec{r})\right|^2$ for the highest occupied molecular orbital (HOMO). The Au tip wave function was simulated by a spherically-symmetric function $\propto \frac{\exp\left[ -k \left( \vec{r}- \vec{r_0}\right)\right]}{k\left|\vec{r}- \vec{r_0}\right|}$ \cite{Tersoff1985} with the decay constant $k=\sqrt{2m\phi}/\hbar$ given by a work function of $\phi=5$\,eV. The integration plane was set at $1.5$\,\AA\  above the center of the molecule. The overlap of the wave functions further depends on the tip height $z$, which is defined as the distance between the center of the s-wave tip and the center of the molecule. Below we will discuss the influence of the tip height in more detail. The data in the main manuscript was simulated at  $z$=8.5\,\AA, which corresponds to a distance of $\sim$ 4.5\,\AA\ before tip--contact formation.

\subsection{Franck-Condon excitation}       

Next, we simulate the set of vibronic resonances within the Franck-Condon picture. In this model, the peak intensity $I_{kn}$ of the $n$th harmonic of a vibrational mode $k$ follows a Poisson distribution:
\begin{equation}
	I_{kn}= e^{-S_{k}} \frac{S^{n}_{k}}{n!},
	\label{eq:Poisson}
\end{equation}

where $S_k$ is the so-called Huang-Rhys factor of the vibrational mode $k$, describing the electron-phonon coupling strength. This factor depends on the relaxation energy $\epsilon_k$ of the vibrational mode upon excitation of the molecule (electron removal for tunneling at negative bias voltage polarity, thereby exciting a vibration in the positive ion resonance):

\begin{equation}
	S_{k} = \frac{\epsilon_k}{\hbar \omega_k}.
	\label{eq:HR}
\end{equation}

\subsection{Franck-Condon parameters from DFT calculations}       

For our simulations, we determined the relaxation energy $\epsilon_k$ and Huang-Rhys factor $S_{k}$ for all vibrational modes from DFT calculations. The method of extracting the parameters follows a description by Wilson, Decius and Cross \cite{Wilson1955}, which has been used in a previous work on BTTT \cite{Krane2018}.
First, we calculated the relaxed structure of the neutral and positively charged molecule in gas phase. This allows to extract the atomic displacements upon charging in mass weighted coordinates $\vec{\eta}_{\alpha}$. Additionally we calculated the vibrational modes and their corresponding normalized mass-weighted atomic displacements $\vec{\l}_{\alpha k}$ for the positively charged molecule.
The relaxation energy can be determined by projecting ($\vec{\eta}_{\alpha}$) onto the vibrational eigenmodes  $\vec{\l}_{\alpha k}$:

\begin{equation}
	\epsilon_{k} = \frac{1}{2} \omega_{k}^{2} (\sum^{N}_{\alpha} \vec{\l}_{\alpha k}\cdot \vec{\eta}_{\alpha} )^{2},
	\label{eq:rel_ener}
\end{equation}

By inserting  this calculated relaxation energy into the equation \ref{eq:HR}, we determined the Huang-Rhys factor $S_k$ for all vibrational modes. We can then simulate the spectra corresponding to the Franck-Condon  excitation ($A_\mathrm{FC}$ in the main text). 
From \ref{eq:Poisson}, we know that the peak intensities for a single vibrational mode $k$ are given by the Poisson-Distribution and are equidistant in energy:

\begin{equation}
 f_k\left(E\right) = e^{-S_k} \sum_{n=0}^{\infty}\ \frac{S_k^n}{n!}\cdot\delta\left(E-n\cdot\hbar\omega_k\right)
 \label{eq:singleModePoisson}
\end{equation}
with $\hbar\omega_k$ its energy.  

Thus, the Franck-Condon spectrum for only one vibrational mode $k$ would be:
\begin{equation}
 A_{FC,k}\left(E\right)=f_k\left(E\right)\ast L\left(E\right)
 \label{eq:singleModeSpec}
\end{equation}
with $L\left(E\right)$ being the peak shape. Here we use a Lorentzian with half-width-at-half-maximum $\gamma$:
\begin{equation}
 L\left(E\right) = \frac{\gamma}{\pi} \frac{1}{\gamma^2+E^2} 
 \label{eq:lorentzian}
\end{equation}

Considering the excitations of several different modes (``progression of progressions'') eq. \ref{eq:singleModeSpec} expands to:
\begin{equation}
  A_{FC}\left(E\right)=f_1\left(E\right)\ast f_2\left(E\right)\ast \cdots \ast f_{m}\left(E\right)\ast L\left(E\right)
  \label{eq:convoluted}
\end{equation}
with a total of $m=3N-6$ different vibrational modes and $N$ being the number of atoms of the molecule.

Eq. \ref{eq:convoluted} can also be rewritten as:
\begin{equation}
	A_{FC} \left( E \right) = e^{-\sum_k S_k}\sum_{n_1=0}^{\infty} \frac{S_1^{n_1}}{n_1!} \,\sum_{n_2=0}^{\infty} \frac{S_2^{n_2}}{n_2!} \,\cdots \,\sum_{n_{m}=0}^\infty 
	\frac{S_m^{n_m}}{n_m!} \cdot L\left(E-\sum_{k=0}^{m} n_{k} \hbar\omega_{k}\right)
\end{equation}

Again using a Lorentzian this yields:
\begin{equation}
	A_{FC} \left( E \right) = e^{-\sum_k S_k}\sum_{n_1=0}^{\infty} \frac{S_1^{n_1}}{n_1!} \,\sum_{n_2=0}^{\infty} \frac{S_2^{n_2}}{n_2!} \,\cdots \,\sum_{n_{m}=0}^\infty 
	\frac{S_m^{n_m}}{n_m!} \frac{1}{\gamma^2+\left(E-\sum\limits_{k=1}^m n_{k} \hbar\omega_{k}\right)^2}	
\end{equation}

In the code used for this publication we take all the modes with $S_{k}>0.001$ and we reduced the number of excitations to three, including also coupled vibronic states, \textit{e.g.} $3\omega_{k}$  or $2\omega_{k}$ + $\omega_{k'}$.



\subsection{Vibration-assisted tunneling}

As discussed in the main text, we suggest to extend the Franck-Condon model by including perturbations of molecular wave functions due to vibrations. Vibration-assisted tunneling can then be regarded as inelastic tunneling within the excited molecular state. 

Inelastic tunneling in the ground state, \textit{i.e.}, neutral molecule, has been experimentally pioneered by W. Ho and colleagues \cite{Stipe1998}, followed by theoretical descriptions by Lorente and Persson \cite{Lorente2000}. 
They considered the change of local density of state due to the perturbation $\delta Q_k\, \partial \Psi_m/\partial Q_k$ of the molecule's wave function and applied Bardeen's theory of tunneling to this perturbed wave function. 
$\delta Q_k=\sqrt{\hbar/2\omega_k}$ is the root mean square displacement of the mass-weighted vibrational normal coordinate $Q_k$ of the mode $k$.
This perturbation of the wave function can be approximated by:

\begin{equation}
 \delta Q_k \frac{\partial \Psi_m}{\partial Q_k} \approx \delta Q_k \frac{\Delta \Psi_m}{\Delta Q_k} \propto \frac{\Psi_{m,k+} - \Psi_{m,k-}}{2}=\frac{\Delta\Psi_{m,k}}{2},
	\label{eq:Approx}
\end{equation}
where $\Psi_{m,k\pm}$  corresponds to the wavefunction for the vibrationally excited molecule, displaced by $\pm \delta Q_k$.
We implemented this in the DFT calculations, by displacing each atom $\alpha$ of the molecule by $\Delta \vec{r}_{\alpha k\pm}= \pm \sqrt{1/m_\alpha}\ \delta Q_k\cdot \vec{l}_{\alpha k}$, with $m_\alpha$ being the atomic mass.
After calculating the wave functions $\Psi_{m,k\pm}$  corresponding to the deformed molecule, we obtained the approximation $\frac{\Delta\Psi_{m,k}}{2}$, according to equation \ref{eq:Approx}.
Thus, the tunneling probability can be calculated for each mode $k$, similar to equation \ref{eq:Bardeen}:

\begin{equation}
	\left|M_{t\Delta_k}(\vec{r}) \right|^2 = \left|\frac{\hbar^2}{4m} \int \Psi_{t} \vec{\nabla} \Delta\Psi_{m,k} - \Delta\Psi_{m,k} \vec{\nabla} \Psi_{t} d\vec{S}\right|^2.
	\label{eq:DeltaBardeen}
\end{equation} 

Note that our experiments address inelastic processes in the charged excited state.
Hence, we calculated the perturbed molecular wave functions and corresponding tunneling matrix elements for the positively charged molecule, which was also used for the calculation of the Franck-Condon excitation.

In analogy to inelastic tunneling in the ground state, the additional contributions to the tunneling matrix element $\left|M_{t\Delta_k}(\vec{r}) \right|^2$ effectively lead to the opening of new tunneling paths, hence, vibration-assisted tunneling. 

To reproduce the full vibrational spectra, we have to add the new tunneling paths associated to the modes $k$ to the simple vibronic spectrum, weighted by the vibration-assisted tunneling matrix element:
\begin{equation}
\begin{aligned}
A_\mathrm{total}(\vec{r},eV) = \left|M_{tm}(\vec{r})\right|^2 \cdot A_\mathrm{FC}(eV) \\
+  \sum_{k} \left|M_{t\Delta_k}(\vec{r})\right|^2 \cdot A_\mathrm{FC}(eV+\hbar \omega_k),
\end{aligned}
	\label{Sp_tot}
\end{equation}

Effectively, this leads to replicas of the Franck-Condon peaks above the threshold energy $E_H +\hbar\omega_k$, with $E_H$ being the resonance energy of the HOMO.

Note that only three parameters are used in the model: the tip-molecule distance (see discussion below), the tip's work function (set to 5\,eV for a gold tip in this work), and the broadening (width of the Lorentzian functions) of the electronic and vibronic states (given by the energy resolution in experiment).

\section{Additional data along BTTT molecule}

\begin{figure}[t]
	\includegraphics[width=\columnwidth]{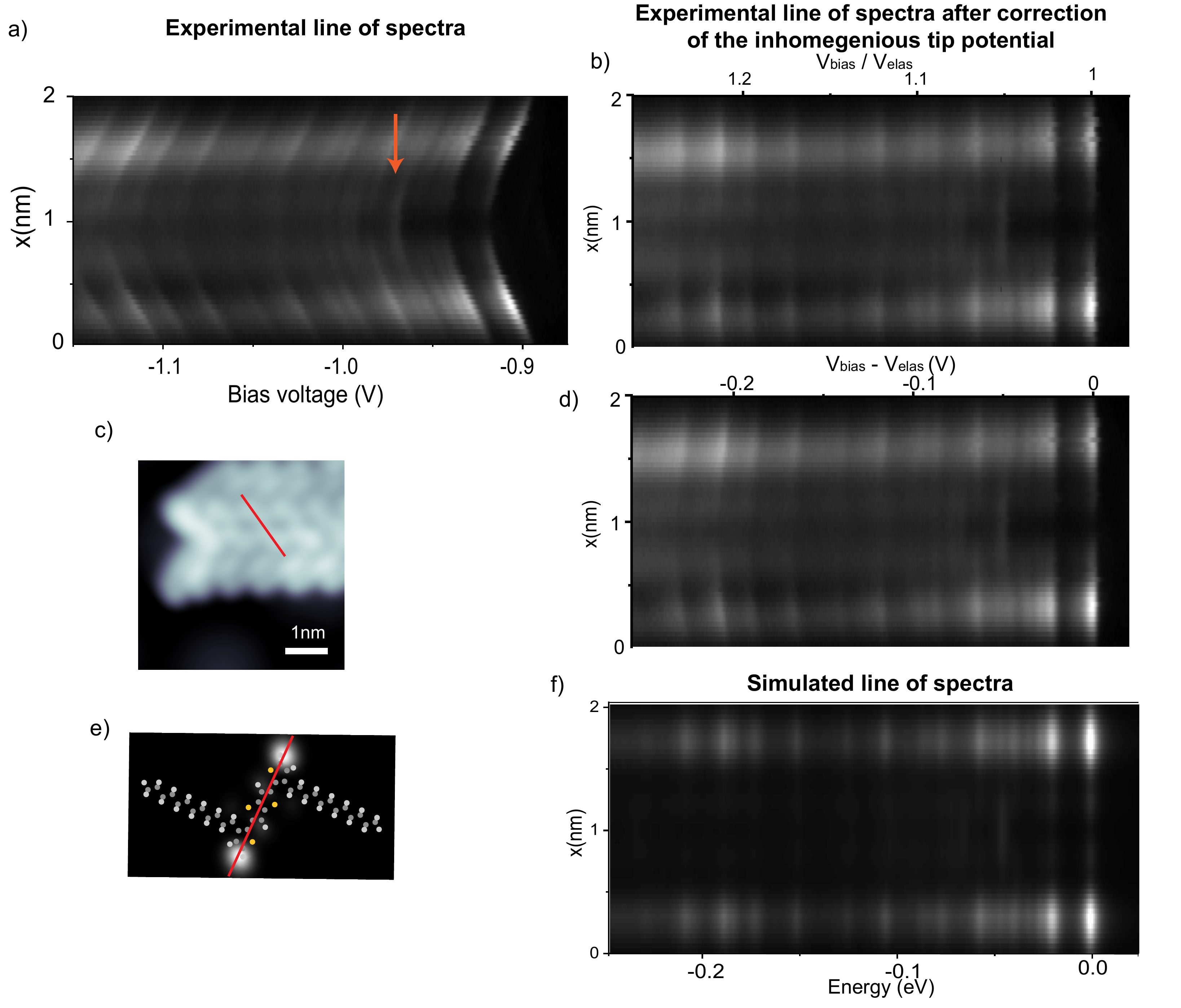}
	\caption{a) 2D plot of 50 \didv spectra recorded along a line over a BTTT molecule (see red line in c)). All spectra are recorded with the same tip height (feedback off). b,d) 2D plot of the same set of data as in a) after accounting for the inhomogeneous tip potential by b) dividing the bias voltage by the voltage of the elastic peak and d) shifting the elastic peak everywhere on the molecule to zero ($V_\mathrm{bias}-V_\mathrm{elas}$). c) STM topography (10\,pA, -850\,mV), of BTTT molecular island. Spectra shown in a,b) are recorded along the red line. e) Simulated conductance map (tunneling matrix element) of the HOMO of the BTTT, with the overlaid molecular model. f) 2D plot of 50 simulated spectra considering the tip position along the red line in d).}
	\label{figS1}
\end{figure} 	

In the main text, we presented \didv spectra recorded at the center and extremity of the thiophene backbone. Here, we show a full set of spectra along the BTTT molecule. Fig. \ref{figS1}a shows a set of 50 spectra recorded along the line shown in Fig. \ref{figS1}c. The spectra were recorded at constant height, $i.e$, the same tip height in all the spectra. Most importantly, the set of vibronic peaks is found all along the molecule with intensity variations as described already in the main text. The orange arrow highlights the most pronounced mode of vibration-assisted tunneling at the center of the molecule. 

The line plot further reveals a smooth parabolic-like shift of the elastic peak and its vibrational sideband by up to $\sim$ 20\,mV to more negative bias voltages in the center of the molecule. This shift was explained in a previous work by the effect of an inhomogeneous electrostatic potential of the tip, which affects the energy levels according to perturbation theory and depending on the tip position \cite{Krane2019}. For direct comparison of the simulated vibrational spectra with experiment we eliminate the inhomogeneous potential in the STM junction in Fig. \ref{figS1}b and d. The most accurate way for this correction is to divide for each spectrum the bias voltage by the bias of the elastic peak (\Figref{figS1}b.). However, in this manner we loose the voltage scaling which makes a direct comparison with the simulations complicated. For this reason, each spectrum in the data set shown in the main text was simply shifted, such that all elastic peaks are found at the same energy. This procedure does not fully account for the inhomogeneous potential of the tip, as we can easily see by the remaining small shift of the vibronic peaks along the molecule (\Figref{figS1}d). However, the remaining inaccuracy amounts only to a few mV, which is almost in the order of our energy resolution and does not change the qualitative results of our comparison between the experiment and the simulation.     

Figure \ref{figS1}e shows a plot of 50 simulated spectra along the molecule (red line in \Figref{figS1}d). Accounting for the voltage drop ($\sim 10\%$) in the \mos, experimental and simulated spectra are in very good agreement, both in vibrational energies and spatial intensity distributions.               

\begin{figure}[h!]
	\includegraphics[width=\columnwidth]{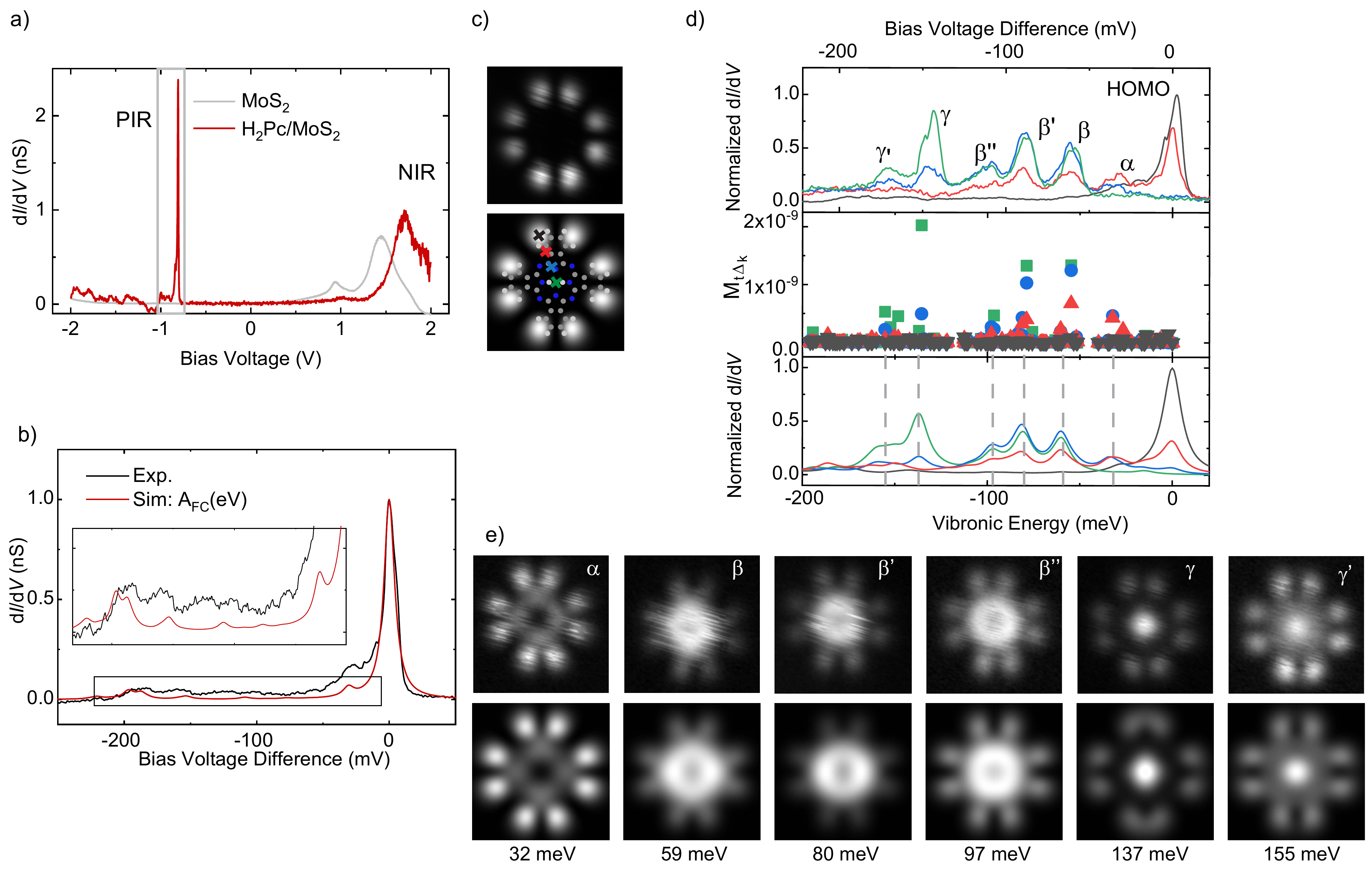}
	\caption{ a) \didv spectra recorded on bare \mos(gray) and on a \HPc (red). Set point: $V$ = 2\,V, $I$ = 300\,pA, $V_\mathrm{mod}$ = 5\,mV. b) Black curve: \didv spectrum in the energy range corresponding to the gray box in a). Set point: $V$ = -1.2\,V, $I$ = 50\,pA $V_\mathrm{mod}$ = 1\,mV. Red curve: simulated Franck-Condon spectrum as described in the theoretical section with Lorentzian functions of 12\,mV FWHM. The energy axis for $A_\mathrm{FC}$ is scaled by a factor of 1.1 to take into account the voltage drop in the \mos, and simplify the comparison with experiment. Inset: zoom of the low intensity part of the spectra (black box). c) Experimental (top) and simulated (bottom) conductance maps of the elastic contribution of the HOMO. d) Top: same experimental \didv spectra than in the main text. middle: Calculated change of the square of the tunneling matrix element $\left|M_{t\Delta_k}(\vec{r})\right|^2$ for all vibrational modes $k$ of \HPc\, for tip positions similar to those shown in Fig. 4 (the same simulated spectra are shown again in the bottom panel) . e) Experimental (top) and simulated (bottom) conductance maps for energies of all inelastic peaks observed and discussed in the main text (See corresponding dashed lines in d)) .}
	\label{figS2}
\end{figure} 

\section{Additional data on \HPc}

Fig. \ref{figS2}a shows a \didv spectrum (red) recorded on a \HPc\, molecule in an energy range spanning over the entire band gap of \mos\ (reference spectrum over the bare \mos\, is also shown for comparison). In the semiconducting gap of \mos\, we observe a sharp positive ion resonance (PIR) of the \HPc. This is at the focus of our discussion in the main text. Moreover, we observe an additional negative ion resonance at $\sim$ 1.75\,V, which can be associated to tunneling through the lowest unoccupied molecular orbital (LUMO) of the \HPc\ \cite{Reecht2019}. Note that this state is not protected by the \mos\, band gap, thus occurring with much broader width due to the much shorter lifetime of the excited state. 

Fig. \ref{figS2}b shows a \didv spectrum around the energy of the PIR, recorded with the tip position over one of the eight lobes of the HOMO. This spectrum exhibits a strong elastic peak, and no intense vibronic replica. Spectra recorded more towards the center of the molecule (Fig. \ref{figS2}d) show intense vibronic features despite of a suppressed elastic peak (see also data in main text). As explained in the main text, we calculated the tunneling matrix elements including vibrational perturbations of the molecular wave function. We observe that many vibrational modes provoke an increase of the tunneling matrix element (Fig. \ref{figS2}d, middle panel). This is due to the presence of a large number of nodal planes in the HOMO, such that symmetry breaking by vibrations plays an important role in this molecule (see more detailed discussion below). Using $\left|M_{t\Delta_k}(\vec{r})\right|^2$, we simulated the total vibrational spectra (as shown in Fig. 4c of the main text), which is in good agreement with the experimental \didv spectra.

Additionally, we show in Fig. \ref{figS2}e the simulated conductance maps for all observed experimental resonances (top panel of \Figref{figS2}e). We observe good agreement for all resonances ($\alpha$, $\beta$ and $\gamma$).

\section{Detailed analysis of modes involved in vibration-assisted tunneling}

\begin{figure}[ht!]
	\includegraphics[width=0.5\columnwidth]{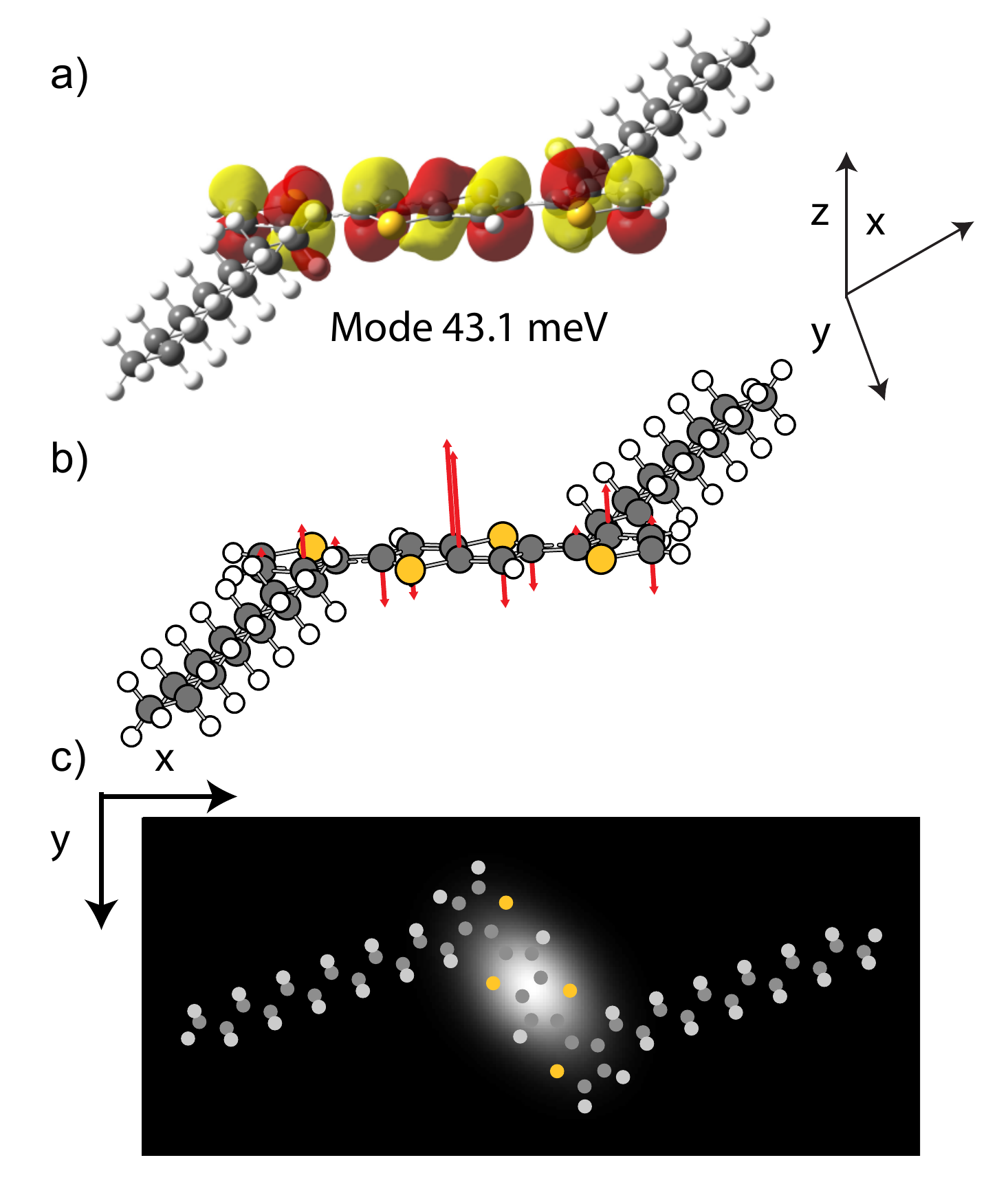}
	\caption{a) Perspective view of the BTTT molecule superimposed with the calculated HOMO wave-function iso-density . b) Visualization of the main vibrational mode involved  in vibration-assisted tunneling, with the same 3D view as in a. The displacement (amplitude and direction) of the atoms upon vibration are represented by red arrows (the motion of H atoms are not shown for clarity). c) 2D map of $\left|M_{t\Delta_k}(\vec{r})\right|^2$ for the vibration described in b.}
	\label{figS3}
\end{figure} 

Considering the exquisite energy resolution in experiment and remarkable agreement with our model of vibration-assisted tunneling, we now aim at characterizing the properties of the vibrational modes, which contribute particularly strongly to this process. To find out about possible selection rules, we describe the character of these modes, and why they participate to the vibration mediated excitation.   

\subsection{BTTT}

Vibration-assisted tunneling in BTTT was only important for one vibrational mode. This mode, with a calculated energy of 43.1\,meV, is represented in \Figref{figS3}b. It describes an out-of-plane stretching motion of the C--C bonds in the thienothiophene backbone (without the alkyl chains). Its influence on the tunneling matrix element can be deduced from the perturbed HOMO wave function. The HOMO exhibits several nodal planes along the thienothiophene backbone (seven lobes, six nodes). The tunneling matrix element is thus smaller in the center of the molecule, where the nodal planes reduce the overlap between tip and HOMO, as compared to the extremities. The vibration at 43.1 meV breaks the symmetry at the center of the molecule, where the C atoms move in opposite out-of-plane directions. Hence, the tunneling matrix element is larger in the center when the perturbation of the HOMO by this vibration is taken into account (see \Figref{figS3}c). The activation of this tunneling matrix element effectively leads to the vibration-assisted tunneling observed in experiment and simulations.


The broken symmetry of the molecular wave function by a vibration with respect to the tip can be regarded as a selection rule for vibration-assisted tunneling. To show its general validity, we also investigated \HPc. We discuss the symmetry arguments at the example of this molecule below.

\subsection{\HPc}

\begin{figure}[h!]
	\includegraphics[width=0.5\columnwidth]{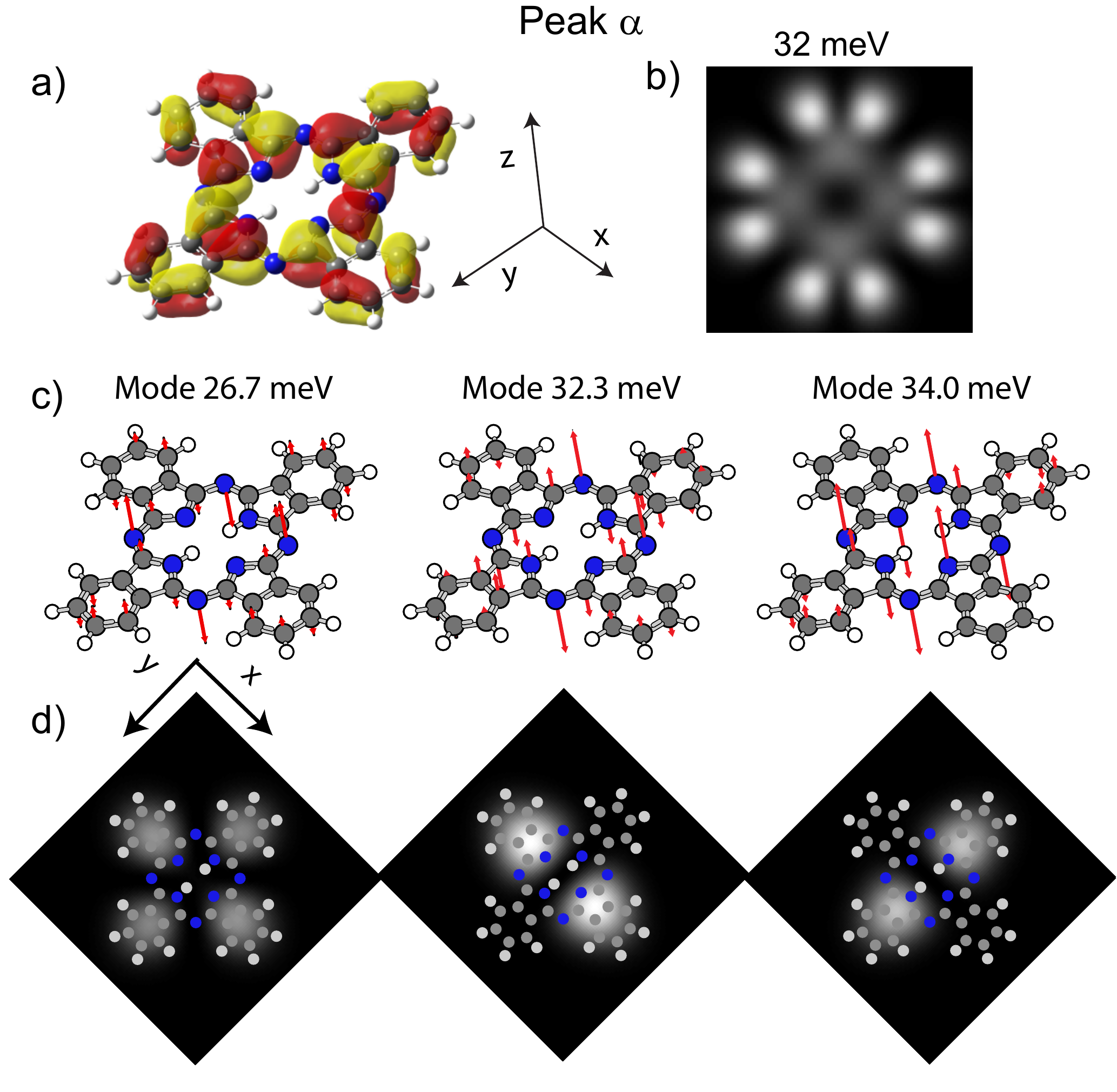}
	\caption{a) \HPc\, molecule superimposed by the calculated iso-density contour of the HOMO wave function. b) Simulated \didv map of resonance $\alpha$. c) Visualization of the vibrational modes involved for the resonance $\alpha$, with the same perspective as in a). The displacement (amplitude and direction) of the atoms upon vibration are represented by red arrows (the motion of H atoms are not represented for clarity). d) 2D map of  $\left|M_{t\Delta_k}(\vec{r})\right|^2$ for the vibrations described in c).}
	\label{figS4a}
\end{figure} 

\begin{figure}[h!]
	\includegraphics[width=\columnwidth]{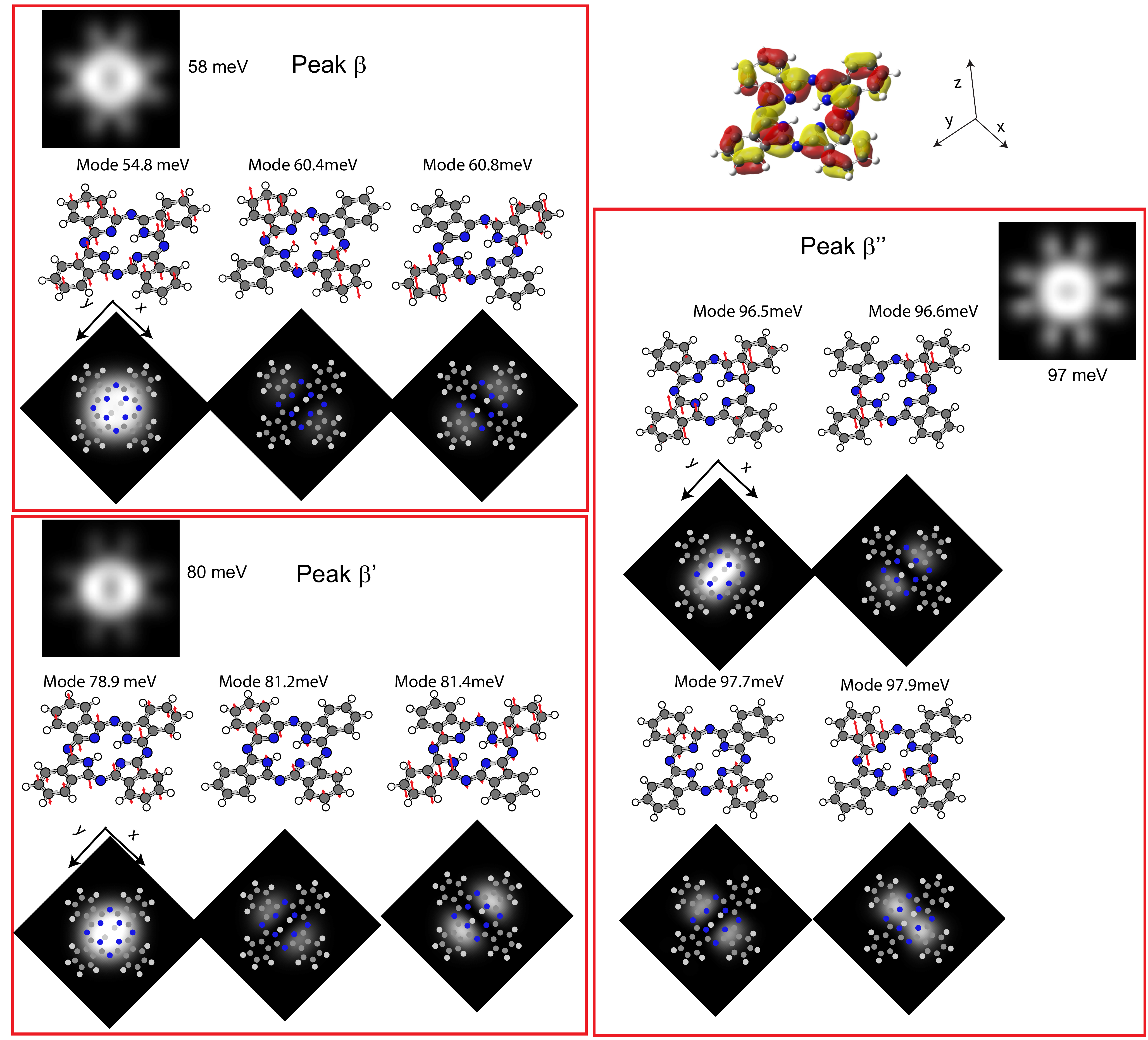}
	\caption{ Visualization for \HPc\, of all the vibrational modes involved for the resonance $\beta$ (top left),$\beta'$ (bottom left) and $\beta''$ (right), their corresponding 2D map  of $\left|M_{t\Delta_k}(\vec{r})\right|^2$ and the simulated map for each resonance. The calculated HOMO's wave-function iso-density is shown in the top right}
	\label{figS4b}
\end{figure}

\begin{figure}[h!]
	\includegraphics[width=0.66\columnwidth]{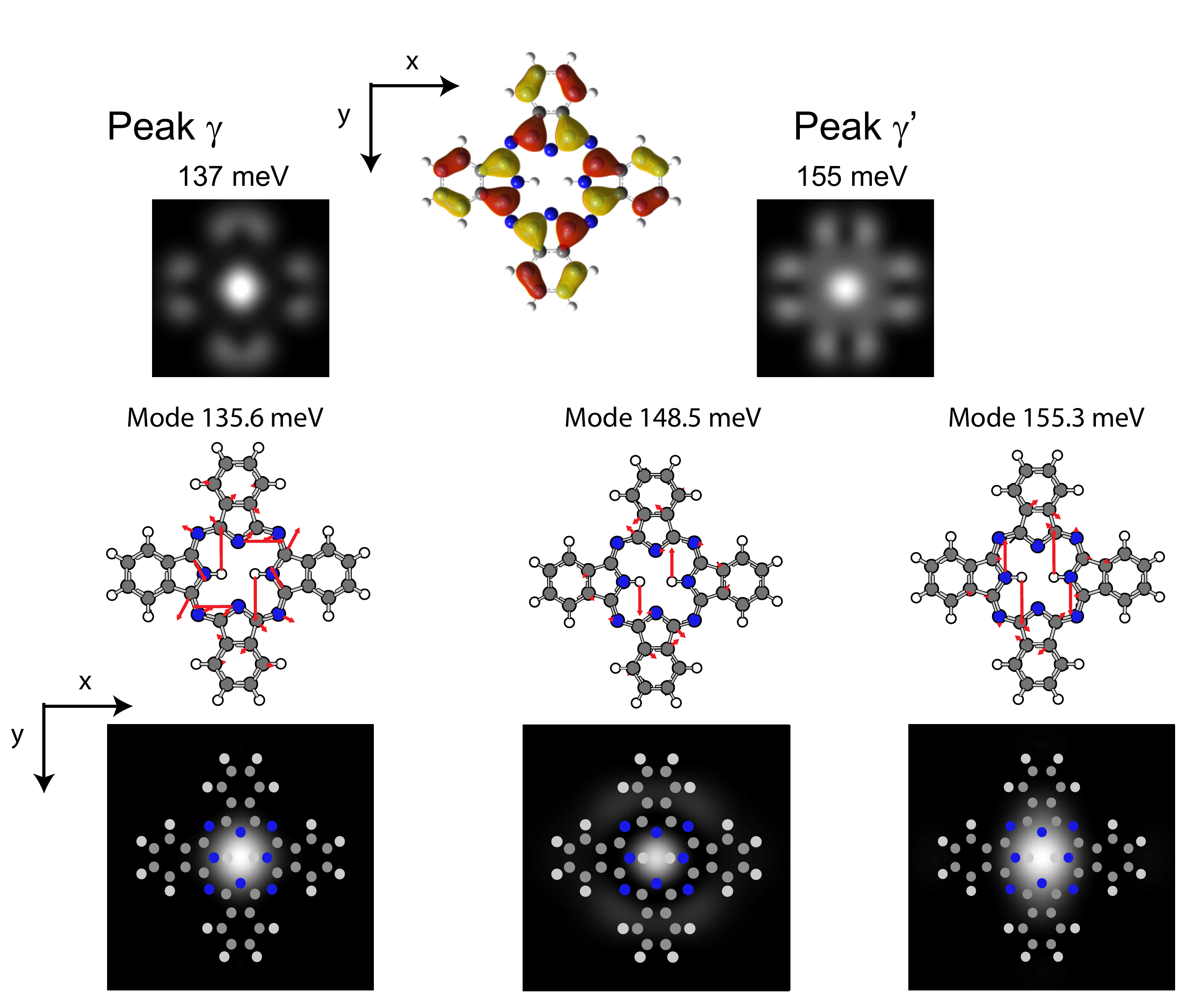}
	\caption{  Visualization for \HPc (top view) of all the vibrational modes involved for the resonance $\gamma$ (left) and $\gamma'$  (right),  their corresponding 2D map  of $\left|M_{t\Delta_k}(\vec{r})\right|^2$ and the simulated map for each resonance. The calculated HOMO's wave-function iso-density is shown in the top (also top view)}
	\label{figS4c}
\end{figure}

As shown in the main manuscript and above, there are several peaks in the \didv spectra that arose from vibration-assisted tunneling in \HPc\ (see also \Figref{figS2}d). These were categorized as $\alpha$, $\beta$ and $\gamma$ according to their symmetric appearance in the experimental \didv maps. We will analyze the properties of the involved modes and their effect on the HOMO now in more detail. 
The vibrational analysis of \HPc\ reveals that each of the experimentally observed resonances consists of several modes, which are close in energy and, thus, cannot be resolved separately in experiment. 
The modes corresponding to $\alpha$, $\beta$ and $\gamma$ are shown in 
\Figref{figS4a}, \Figref{figS4b}, and \Figref{figS4c}, respectively. 


Resonance $\alpha$ arises from three modes at 26.7, 32.3 and 34\,meV. These modes describe out-of-plane stretching modes of the C--N bonds of the macrocycle (\Figref{figS4a}c). The first one breaks the symmetry of the wave function along the nodal plane perpendicular and parallel to the inner H atoms. The second and third mode break the symmetry perpendicular (32.3\,meV) and parallel (34\,meV) to the nodal plane along the H atoms, respectively. The tunneling matrix elements associated to the vibration-induced perturbations of the wave functions are shown \Figref{figS4a}d. The sum of the three vibration-assisted and elastic tunneling contributions yield the map in \Figref{figS4a}b.


The mode analysis for the $\beta$ resonances is shown \Figref{figS4b}. Again, $\beta$, $\beta'$ and $\beta''$ arise from several energetically close-lying vibrations. Common to all vibrations is their out-of-plane character. They describe molecular distortions due to out-of-plane stretching of C--C bonds of the macrocycle and/or the isoindoline moieties (see \Figref{figS4b}). Moreover, all modes share their anti-symmetric character with respect to the perpendicular (modes at 60.4, 81.2, 97.7\,meV) or the parallel (modes at 60.8, 81.7, 96.6\,meV) direction of the inner H atoms or simultaneously both directions (54.8, 78.9, 96.5 and 97.9\,meV). Furthermore, the modes at 54.8 and 78.9\,mV  are also asymmetric with respect to the node along the diagonal of the macrocycle. Hence, multiple modes delocalized along the macrocycle break the symmetry at the numerous nodal planes of the HOMO wave function. The vibration-assisted tunneling matrix element $\left|M_{t\Delta_k}(\vec{r})\right|^2$ (\Figref{figS4b}) is therefore sizable along the entire macrocycle.

The $\gamma$ resonances originate from in-plane vibrations with an asymmetric deformation of the inner macrocycle, mainly discernible by the motion of the inner H and N atoms (see scheme in \Figref{figS4c}). These modes break the symmetry relative to the center of the molecule and crucially affect the overlap with an s-wave tip, expressed by a large $\left|M_{t\Delta_k}(\vec{r})\right|^2$.             

\section {Height dependence of vibration-assisted tunneling} 

\begin{figure}[h]
	\includegraphics[width=\columnwidth]{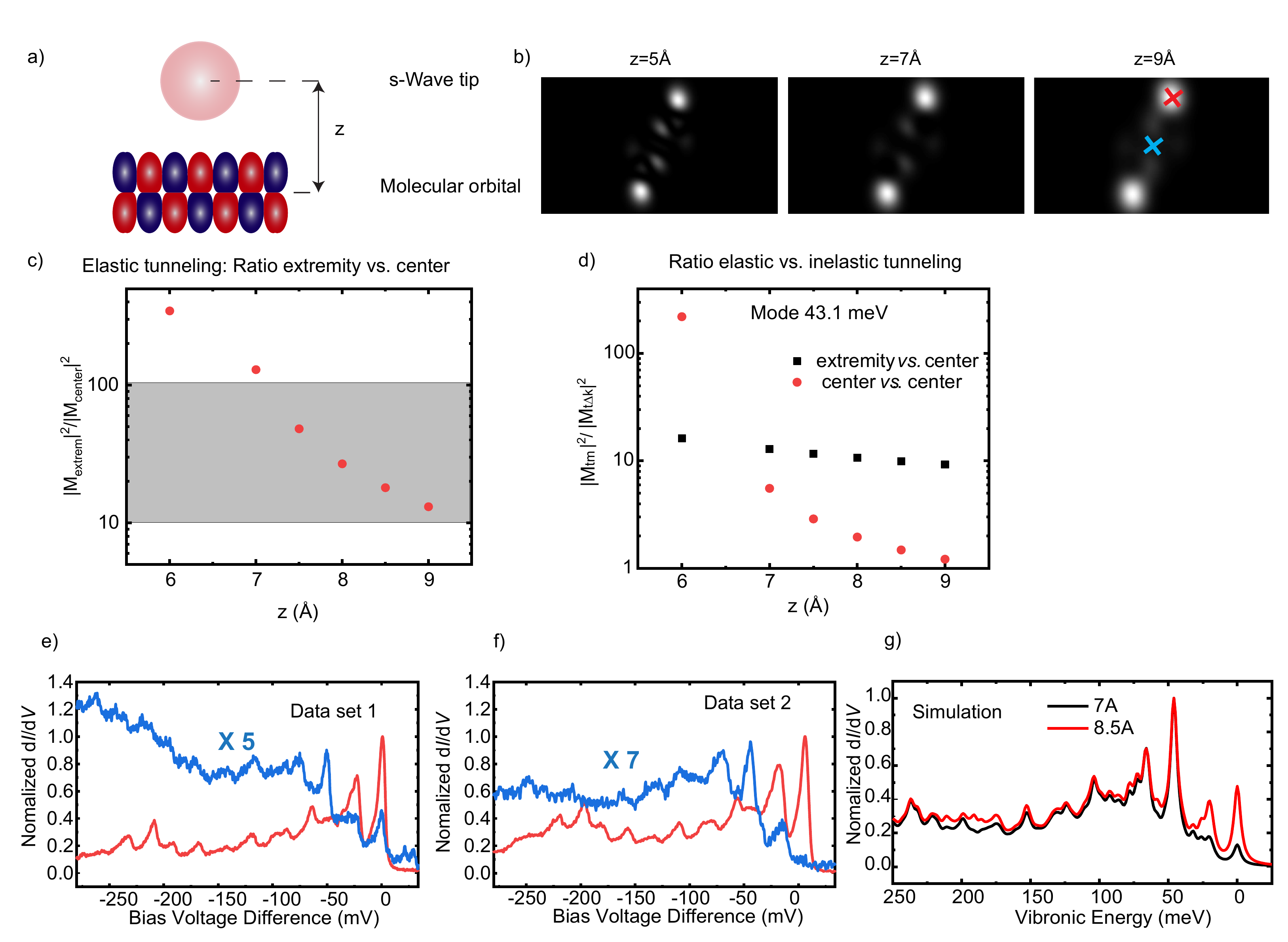}
	\caption{ a) Sketch of the wave functions of tip and molecule, with the definition of $z$ as used in the model. b) Simulated conductance map for the HOMO of BTTT for three different values of $z$. c) $z$ dependence of the ratio of the tunneling matrix element $\left|M_{tm}(\vec{r})\right|^2$ at the extremity and center of the molecule (for positions see crosses in b). d) $z$ dependence of the ratio of the tunneling matrix element between the elastic $\left|M_{tm}(\vec{r})\right|^2$ (tip above the extremity (black) and the center (red)) and the inelastic $\left|M_{t\Delta_k}(\vec{r})\right|^2$ (tip above the center). e) and f) experimental \didv spectra for two different molecules with the tip above the extremity (red) and the center (blue) (feedback open with tip above the extremity at -1.2\,V and (e) 300\,pA or (f) 250\,pA  ). g) simulated \didv spectra for the tip above the center of the molecule for two different values of $z$.}
	\label{figS5}
\end{figure}

As explained in the description of the model, the tip height (or tip-molecule distance), is one of the free parameters in the calculations of the tunneling matrix elements. In this section we discuss how this parameter influences the relative intensity of vibration-assisted tunneling, and therefore the comparison with the experiment.

To start with, we need to clarify that the tip height $z$ in the calculations of the tunneling matrix element is defined as the distance between the center of the s-wave tip and the center of the molecule (see \Figref{figS5}a). Therefore the values of $z$ used in this section cannot be directly compared with the tip--molecule distance considered in STM experiments, which is commonly referred to as the tip distance  to ``contact'' of the molecule. 

\Figref{figS5}b shows three simulated conductance maps for the HOMO of the BTTT at three different tip heights (5, 7 and 9\, \AA). At larger $z$ the structure of the nodal planes is washed out due to the extent of the wave functions in vacuum. A similar effective reduction of spatial resolution may also result from a broad tip apex. While we consider an s-wave tip shape, the  tip apex  may consist of one or several atoms in reality. As we neither know the exact tip height, nor the tip shape, the parameter $z$ implicitly accounts for the interplay of both of these. Thus, adjusting the tip height $z$ is also a way to simulate different tip shapes, with a smaller (higher) $z$ associated to a sharper (blunter) tip apex. 

The effect of $z$ in the model can be observed in the case of BTTT by comparing the elastic contribution with the tip above the extremity and in the center of the molecule (same positions as in the main text, indicated by crosses in \Figref{figS5}b). \Figref{figS5}c shows the ratio of the tunneling matrix elements $\left|M_{tm}(\vec{r})\right|^2$ at these two positions as a function of $z$. The ratio decreases continuously from $\sim 300$ at $z=6$\, \AA\  to $\sim$ 10 for $z=9$\, \AA. This variation results from the effect of smoothing of the wave function overlap above the molecule as described above. 
This range is in agreement with experiment, where we observe different intensity ratios (from 10 to 100, indicated by the gray box of \Figref{figS5}c) of the elastic peaks at the BTTT's extremity and center with different tips - with nominally the same tip height.  An example of the variation of the spectra with different tips is shown in 
\Figref{figS5}e and f. The elastic peak in the center is hardly observable in the second data set. As the conductance of the elastic peaks is the same at the extremity in both cases, the change in ratio does not indicate a different tip height, but another tip shape. To show that this may indeed be the case, we calculate the ratio of the elastic ($\left|M_{tm}(\vec{r})\right|^2$)  and inelastic ($\left|M_{t\Delta_k}(\vec{r})\right|^2$) contribution as a function of $z$ (\Figref{figS5}d). The ratios observed in experiment correspond to $z=8.5$\, \AA\ for the data set 1 and $z=7$\, \AA\ for the data set 2. The change in tip shape may thus be captured in the simulations by a variation of the free parameter $z$.


\begin{figure}[h]
	\includegraphics[width=0.4\columnwidth]{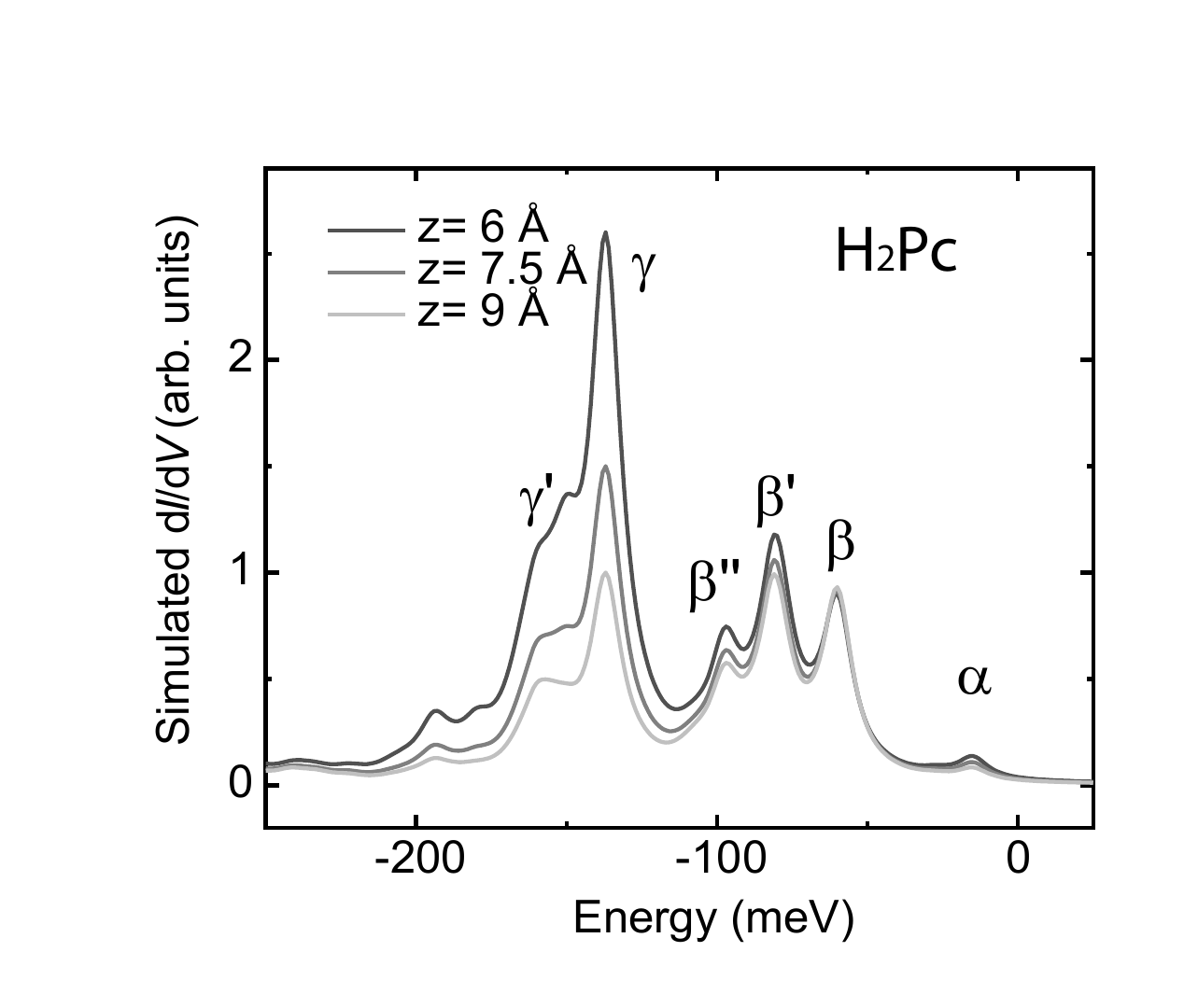}
	\caption{ Simulated spectra of the \HPc, considering the tip above the center of the molecule,  for three different values of $z$. Spectra are normalized in intensity to the resonance $\beta$.}
	\label{figS5b}
\end{figure}

A last remark on the height dependence of the model refers to the relative efficiency between different modes of the vibration-assisted tunneling. \Figref{figS5b} shows, in the case of \HPc, a set of simulated spectra for three different values of $z$ (all other parameters are similar), considering the tip above the center of the molecule. For better comparison, the spectra are normalized on the $\beta$\, resonance. While the $\alpha$\, and $\beta$ peaks keep almost the same relative intensities, the $\gamma$ peaks vary  strongly. This indicates that at this tip position, vibration-assisted tunneling due to the modes involved in the resonances $\gamma$ decay faster with $z$ than for the other modes. As we discussed before, the modes involved for $\gamma$ and $\gamma$' are in-plane vibrations whereas out-of-plane modes dominate the other resonances. 
Intuitively, we can easily understand that in-plane and out-of-plane perturbations of the wave function react differently to tip approach. Unfortunately, for stability reasons (molecules diffuse easily upon tip approach), we do not have experimental confirmation of this effect.


\begin{thebibliography}{37}
\bibitem [{not({\natexlab{a}})}]{note}%
  \BibitemOpen
  \href@noop {} {} \bibinfo {note} {For the calculation of
  the tunneling matrix element, the tip-molecule distance is defined between
  the center of the s-type wave function and the center of the molecule. This
  value is therefore larger than the distance of the tip to contact formation
  with the molecule. For more details see SI.}\BibitemShut {Stop}%
\bibitem [{not({\natexlab{c}})}]{note3}%
  \BibitemOpen
  \href@noop {} {} \bibinfo {note} {If this peak was a
  second harmonic peak with higher intensity than the elastic peak, the
  Franck-Condon principle would predict an additional higher harmonic at $\sim
  100$\,meV, which is not to be found.}\BibitemShut {Stop}%
\end{thebibliography}
\end{document}